# X-RAY EXPERIMENT PROVIDES A WAY TO REVEAL THE DISTINCTION BETWEEN DISCRETE AND CONTINUOUS CONFORMATION OF MYOSIN HEAD

E. V. Rosenfeld

Institute of Metal Physics, Ural Branch of Russian Academy of Sciences
S Kovalevskoy 18, Yekaterinburg, 620990  Russia
e-mail: rosenfeld@imp.uran.ru

*Abstract*

The corner stone of the classical model after Huxley and Simmons, which has been serving as a basis for the theory of muscle contraction for already four decades, is supposition that a myosin head can reside only in several discrete states and irregularly jumps from one state to another. Until now, it has not been found a way to experimentally verify this supposition although confirmation or refutation of the existence of discrete states is crucial for the solution of myosin motor problem. Here I show that a set of equal myosin heads arranged equidistantly along an actin filament produce X-ray pattern which varies with the type of conformation. If the lever arms of all myosin heads reside in one and the same position (continuous conformation), all the heads have the same form-factor and equally scatter electromagnetic wave. In this case, only the geometric factor associated with a spatial ordering of the heads will determine the X-ray pattern. The situation changes if the average lever arm position is the same, but inherently every head can reside only in several diverse discrete states, hopping irregularly from one to another. In this case, the form-factors corresponding to distinct states are dissimilar, and the  increments in phases of  X-rays scattered by different heads are different as well. Inasmuch as every quantum of radiation interacts with the heads residing in different states, this results in additional interference and some peaks in the X-ray pattern should slacken or even extinct as compared with the pattern from the heads with the continuous-type conformation. The formulas describing both cases are compared in this article. In general, the distinction between X-ray patterns is insignificant, but they could be appreciably different at some stages of conformation process (respective lever arm position depends on the type of discrete model). Consequently, one can with luck attempt to find out this difference using a high-sensitive equipment.

**Introduction**

Myosin head (MH) is an elementary force generator (myosin motor) with a characteristic size of the order of 10 nm, which up to now has made it virtually impossible to directly observe what is happening inside it. Yet, short-wave radiation, when scattering on a system of uniformly oriented MHs which are in one and the same state and spatially ordered with a periodicity, can provide valuable information about changes in their internal structure upon conformation. To drive most of MHs into an approximately equal conformation and spatially order them is possible via realization for the muscle fiber of the isometric contraction state. Following this step, a rather fast change in the fiber length leads to a synchronous conformation change of the attached cross-bridges. This method is in essence an analog of the one used in pioneer works (Podolsky 1960; Huxley and Simmons 1971). However, for the last decades, the technique of examining MH with the use of small-angle X-ray scattering has been significantly improved (Irving et al. 1992; Irving et al. 2000; Juanhuix et al. 2001; Piazzesi et al. 2002; Reconditi et al. 2004; Huxley et al. 2006 a,b; (Reconditi 2006); Reconditi et al. 2011; Brunello et al. 2014). Since, when using powerful synchrotron radiation, the minimal time of exposure for a single X-ray pattern decreases to 100 μs and more than this (Dobbie et al. 1998), a change in the MH state for several ms can be traced in much detail.

The most thorough examination of the MH conformation in the course of working stroke (process of force generation) is necessary for ascertainment of the principle of myosin motor operation. Therefore, a crucial role is ascribed to the adequate interpretation of X-ray patterns, which by no means provide unambiguous information on the object. Although a few vectors of the space lattice whose sites are occupied by scattering elements (in our case - MHs) can be more or less unambiguously determined from the positions of the fundamental reflections, the internal structure of these elements is much more difficult to ascertain. It exerts an effect only on the form of peaks and their amplitude ratio, while in the absence of a strict spatial order (which is quite typical of muscle fibers), the reflections are greatly smeared. This increases uncertainty in the initial data that are used for the determination of the internal structure of MHs in the interpretation of X-ray patterns. Therefore, only based on a clear structural model of the process of MH conformation change, one can match different values of its parameters with different stages of this process.

It is commonly accepted that a MH consists of two relatively rigid elements fastened to each other – motor domain and lever arm, and in the process of conformation there takes place a turn of the lever arm (see the excellent review by Sweeney and Houdusse (2010) and the Refs. therein). However, there exist two principally different models of the very process of rotation.

The theory after Huxley and Simmons (1971) and, correspondingly, a lot of models patterned after it are based on the supposition on the existence of discrete states for MH. Such supposition is introduced in order that to explain strange features of relaxation arisen after a sharp change in the fiber length. Theories of such type present a convenient way of formally describing results of quick release/stretch experiments. However, they absolutely do not take into consideration the mechanism of emergence inside the MH of a force that stretches an elastic element EE connecting the MH with the myosin rod and causes sliding of filaments (see Fig.1).

The first theory in which the conformation is treated as a continuous change of the MH state was also proposed by A.F. Huxley (1957). Since in such theories no supposition on the preference of one of the directions of rotation due to thermal fluctuations, which contradicts the second law of thermodynamics (Feynman et al. 1967), is employed, they cannot explain the anisotropy of the relaxation processes in the step length changes experiments. However, a continuous conformation allows one to explain the force-velocity curve proposed by Hill (1938), and it can be traceable to the action of Coulomb's forces arising after the ATP hydrolysis (Rosenfeld 2012; Rosenfeld and Günther 2014).

Regardless of the type of conformation, one can assume, as is customary when the isometric contraction state is considered, that the distance between the points of attachment of adjacent MHs to actin filament is approximately constant. Nevertheless, this does not mean that X-ray diffraction patterns should be identical in the different-type models. In the case of continuous conformation, all the MHs are in the same state and equally scatter X-rays. However, in the case of conformation of the Huxley-and-Simmons (1971) type, the heads occupying the positions located periodically along actin fiber reside at any time in different states and, therefore, scatter X-rays in a number of ways. This situation, which arises when passing from an individual substance to a substitutional alloy, is well known in the solid-state physics. The more complicated problem when all the atoms are equivalent, but everyone can take one of several different positions in its unit cell (hydrogen atom leaping along the interstitial sites; compare it with a lever arm which can be oriented diversely) was for the first time considered by Kassan-Ogly and Naish (1986). Based on these results, it is shown below that two types of myosin head models predict different changes in the X-ray pattern in the process of conformation change. It is not a priori evident if the accuracy of the modern experimental techniques is sufficient for the detection of these differences; yet, a principle possibility arises to clarify whether actually the lever arm can smoothly rotate or only a few discrete states are allowed and a seemingly continuous rotation is controlled by fast jumps over these states upon changing probabilities of their occupation.

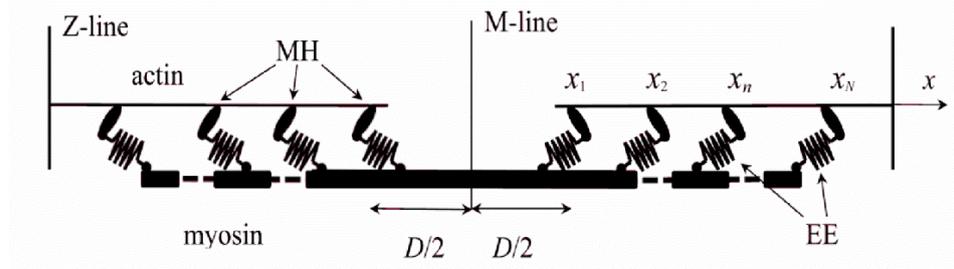

**Figure 1.** The simplest model of sarcomere. *N* myosin heads MH per half sarcomere are fastened on myosin rod with elastic elements EE and equidistantly attached to actin filament at the points $x_n = \dfrac{D}{2} + (n-1)d$. The origin of the abscise axis *x* coincides with the M-line; the width of the bare zone, i.e. the zone free of MHs is equal to *D*.

### Results

Let us consider the simplest model of sarcomere depicted in Fig.1, see also (Piazzesi et al. 2002; Reconditi et al. 2004). Suppose that *N* unit MHs are attached to an actin filament at points $x_n = \dfrac{D}{2} + (n-1)d$ in one part of the sarcomere and other *N* units, in the other part at points $x_{-n} = -x_n$, $n = 1, 2, \ldots N$. Then, the amplitude of elastic scattering of electromagnetic radiation is equal to

$$A\left(\lambda, \{s\}\right) \sim \frac{1}{2N} \sum_{\substack{n=-N \\ (n \neq 0)}}^{N} F\left(\lambda, s_n\right) \exp\left(2\pi i \frac{x_n}{\lambda}\right) \tag{1}$$

Here, $\lambda$ is the wave length corresponding to *x*-component of the scattering wave vector (the difference between the wave vectors of the scattered and incident waves) and $s_n$ is the internal coordinate of the *n*-th MH, which determines its state; distribution of $s_n$ over MHs is denoted as $\{s\}$. As the internal coordinate, any coordinate can be chosen that determines the position of the lever arm with respect to the motor domain. The function *F* in Eq. 1 specifies the form factor of a MH in the state *s*:

$$F\left(\lambda, s\right) = \int \rho_s\left(x, y, z\right) \exp\left(2\pi i \frac{x}{\lambda}\right) dx\,dy\,dz \Rightarrow \left|F\left(\lambda, s\right)\right| \exp\left\{2\pi i \frac{a\left(\lambda, s\right)}{\lambda}\right\} \tag{2}$$

Here, $\rho_s$ is the density of distribution of X-ray scatters (substantially electrons) inside the MH resting in the state *s*, and to eliminate the dimension-oriented problems an arrow is used instead of equal mark. The origin of coordinate in Eq. 2 is chosen at the point of the MH attachment (point $x_n$ for the *n*-s MH), so that the variable *a* which determines the form-factor phase is equal to displacement (in the course of lever arm rotation) of the "scattering center" of the MH relative to this point.

Thus, the form-factor specifies the complex amplitude of radiation scattered by a single MH, and the sum in Eq. 1 gives the result of interference of X-rays scattered by all MHs. Since the time of interaction of a single X-ray photon with muscle fiber is negligibly small in comparison with the time of changing the system state, the intensity $I$ of its scattering is equal to

$$I\left(\lambda,\{s\}\right) = \left|A\left(\lambda,\{s\}\right)\right|^2 \qquad (3)$$

For the exposure period $\Delta t$, a plenty of quanta are scattered and their interaction with the counter is equivalent to the summing-up of intensities. If the system state is not changed for the time of interaction, the average value $\left\langle I\left(\lambda,\{s\}\right)\right\rangle = I\left(\lambda,\{s\}\right)$. For instance, if one observes the continuous conformation after a length step change (characteristic time about 1 ms) and shoots a "film" at 50 μs intervals, the natural preposition is that in every picture all the MHs are in a well-defined state $s$, one and the same for all of them, but it varies from picture to picture $s = s(t)$.

Yet, if for the period $\Delta t$ every MH has the time to reside in several states, the X-ray pattern is a result of summing-up of intensities for all the states of the system taken with the corresponding statistical weights $P\left(\{s\}\right)$:

$$\left\langle I(\lambda)\right\rangle = \sum_{\{s\}} P\left(\{s\}\right)\left|A\left(\lambda,\{s\}\right)\right|^2. \qquad (4)$$

The same formula should be used if $\Delta t$ is not large in comparison with the time of the single jump of a MH between its states but we observe many independent systems simultaneously (a lot of equal sarcomeres are illuminated by X-ray).

### *Continuous lever arm rotation*

If all cross-bridges are in equal states $s$ (*e.g.* in isometric contraction state), then the form-factors Eq. 2 of the MHs from the left and right half of the sarcomere in Fig.1 are complexly conjugated $F_n = F_{-n}^*$ (we drop index $s$ and the symbol * denotes complex conjugation). Now, from Eq. 1 we obtain

$$A\left(\lambda,L,a\right) \sim \left|F\right|\cos\left(\pi\frac{2a+L}{\lambda}\right)\frac{\sin\left(\pi N d/\lambda\right)}{N\sin\left(\pi d/\lambda\right)}, \quad L = D + (N-1)d \qquad (5)$$

Here, in accordance with designations (Koubassova et al. 2009), $L$ is the distance between the attachment points $x_n - x_{n-N-1}$ of equivalent MHs from two halves of sarcomere, and the addendum $2a$ arises because of an additional symmetrical divergent displacements of their "scattering centers", see (2). The intensity of radiation scattered by each of two "diffraction grids" in the right and left parts of sarcomere is defined by the Laue function

$$I_{Laue}(\lambda) = \frac{1}{N^2} \left| \sum_{n=1}^{N} \exp\left( 2\pi i \frac{nd}{\lambda} \right) \right|^2 = \left\{ \frac{\sin\left( \pi N \frac{d}{\lambda} \right)}{N \sin\left( \pi \frac{d}{\lambda} \right)} \right\}^2 . \tag{6}$$

All of $N$ MHs from each of the sarcomere parts scatter waves in phase, if $d = \lambda \times \text{integer}$. It is just these values of $\lambda$ at which $I_{Laue}(\lambda)$ is maximal. The result of interference of rays going from both parts of sarcomere in Eq. 5 is controlled by the «optical length difference» $2a + L$. If it is equal to an integer/half-integer number of wave lengths, the rays amplify/dampen each other:

$$I(\lambda, L, a) = \left| F(\lambda) \right|^2 \cos^2\left( \pi \frac{2a(\lambda) + L}{\lambda} \right) I_{Laue}(\lambda) \tag{7}$$

Using this formula, it is convenient, for example, to trace changes in the position and internal state of MHs via observing changes in the structure of reflection M3 at $\lambda = d$ (see *e.g.* Reconditi *et al* (2006) and the references therein). This peak is split into two virtually symmetrical peaks at $D + 2a = d \times \left( \text{integer} + \frac{1}{2} \right)$, since cosine in Eq. 7 vanishes in its center. With changing $D + 2a$, one of these peaks decreases whereas the other grows, and, gradually, becomes the only one peak whose height reaches maximum at $D + 2a = d \times \text{integer}$. Truly speaking, while analyzing, there arise some complications connected with the fact that the shift of MHs together with filaments (change of $D$) gives rise to the same change in the X-ray pattern as does the MHs conformation *i.e.* change of $a$ (see Rosenfeld 2014).

It is just a formula of the type Eq. 7 that must be used if it is assumed that the lever arm can take any (but one and the same for all the MHs) position upon rotation. However, these formulas are used as well for comparison of the experimental and computational results, employing models of Huxley and Simmons (1971)-type, see *e.g.* (Irving et al. 2000; Koubassova et al. 2009). In these works, after calculation of occupation probabilities for all discrete states corresponding to different orientations of the lever arm, an averaged position of the lever arm is calculated. In the frame of Huxley and Simmons (1971)-type models, the lever arm never stays in this position; this is solely the way of averaging which can be used when determining the average length of the EE. Yet, further, the corresponding form-factor is calculated to be substituted into Eq. 7.

Such method of averaging cannot be applied to calculation of interference effects if at any moment MHs can be found in different states with certain probabilities. In accordance with Eq. 4, it is squared modules of scattering amplitudes, *i.e.,* bilinear combinations of form-factors, that must be averaged rather than internal coordinates of MHs. Besides, the disordering caused by transitions between states leads to lowering the reflection intensities and to appearance of diffusion scattering (Kassan-Ogly and Naish 1986). These problems are discussed in what follows.

### *Discrete lever-arm states*

If MHs can stay solely in definite discrete states $s^{(1)}, s^{(2)}, s^{(3)},...$ and, with time, pass from one state to another, then, the resulting X-ray pattern will depend on the exposure $\Delta t$. Of special interest will then be the case when $\Delta t$ is so large that each MH has the time to take any of these states. In this case, the probability for a MH to stay in state $s^{(\nu)}$ (it is equal to the ratio of the total time of residing in this state to $\Delta t$) will be denoted as $p_\nu$. These very probabilities are determined via calculations within the corresponding discrete models. $p_\nu$ are normalized: $\sum p_\nu = 1$, and do not depend on the MH's position $n$ if all of the heads are in equal conditions. Moreover, the probability for the system to be in state $\{s\} = \left\{ s_{-N}^{(\nu_{-N})},...,s_N^{(\nu_N)} \right\}$ is equal to the product $P\{s\} = \prod_{\substack{n=-N \\ (n \neq 0)}}^{N} p_{\nu_n}$. Therefore, substituting into Eq. 4 the explicit form of the scattering amplitude Eq. 1, we have

$$\langle I \rangle = \frac{1}{4N^2} \left\{ \sum_{\substack{m=-N \\ (m \neq 0)}}^{N} \left[ \left( \sum_\mu p_\mu F_\mu \exp\left( 2\pi i \frac{x_m}{\lambda} \right) \right) \times \sum_{\substack{n=-N \\ (n \neq 0) \\ (n \neq m)}}^{N} \left( \sum_\nu p_\nu F_\nu^* \exp\left( -2\pi i \frac{x_n}{\lambda} \right) \right) \right] + \sum_{\substack{m=-N \\ (m \neq 0)}}^{N} \sum_\mu p_\mu |F_\mu|^2 \right\} \tag{8}$$

The summands in the first sum are mutually complex conjugate, and each term is the product of the parameters related to two *distinct* MHs with numbers $m$ and $n$ residing in the states $\mu_m \to \mu$ and $\nu_n \to \nu$ with probabilities $p_\mu$ and $p_\nu$ respectively. The summands with $m=n$ of the product $A\left( \lambda, \{s\} \right) A^*\left( \lambda, \{s\} \right)$ involve the scattering amplitudes related to only one MH, obtain the single probability of state $p_\mu$ (see Eq. 4) and compose the last sum in Eq. 8.

Removing the restriction $n \neq m$ from the first sums in Eq. 8, we obtain the additional terms with $m=n$ and all possible $\mu, \nu$. Subtracting simultaneously the same terms from the last sum, we can reduce Eq. 8 to the form

$$\langle I(\lambda) \rangle = |\Phi(\lambda)|^2 \cos^2\left( \pi \frac{2\varphi(\lambda) + L}{\lambda} \right) I_{Laue}(\lambda) + \frac{1}{2N} \left\{ \sum_\mu p_\mu |F_\mu(\lambda)|^2 - |\Phi(\lambda)|^2 \right\} \tag{9}$$

This formula must replace Eq. 7 for the case of MHs with discrete states and the sum

$$\Phi(\lambda) = \sum_\mu p_\mu F_\mu(\lambda) = |\Phi(\lambda)| \exp\left(2\pi i \, {}^{\varphi(\lambda)}\!/\!_\lambda\right) \qquad (10)$$

substitutes for form-factor $F$ Eq. 2. The first summand in Eq. 9 represents the dependence on $\lambda$ of the peak intensity of coherent scattering, whereas the second summand, that of diffuse scattering (Kassan-Ogly and Naish 1986).

### Discussion

To clarify the foregoing let us consider a few simple examples and the physical meaning of the results.

#### Simple examples

To illustrate the distinction between Eq. 7 and Eq. 9, let us consider the simplest case when in the course of MH conformation the modulus of its form-factor doesn't change:

$$|F(\lambda, s)| = |F| = const \qquad (11)$$

whilst its phase changes according to Eq. 2. Let's assume for the beginning that in a discrete model MH can reside only in two states and form-factors

$$F_{1,2}(\lambda) = |F| \exp\left(\mp 2\pi i \, {}^a\!/\!_\lambda\right). \qquad (12)$$

correspond to these states. This means that in the first state the scattering center is displaced by $a$ toward the M-line relative to the point of the MH attachment while in the second state it is displaced by $a$ in the opposite direction. In the course of conformation, the probabilities $p_1$ and $p_2$ vary and this results in the displacement of two imagined objects: "averaged position" of the lever arm and the corresponding "averaged position" of the scattering center. However, X-rays are scattered by real rather than imaginary objects. Consequently, the formulas (7) and (9) give the same results if $p_1 = 1$, $p_2 = 0$ (lever arm occupies position $+a$) or $p_1 = 0$, $p_2 = 1$ (lever arm occupies position $-a$) but situation changes if, say, $p_1 = p_2 = {}^1\!/\!_2$.

In this case the "averaged position" of the lever arm coincides with the MH's attachment point and using the method described in (Irving et al. 2000; Koubassova et al. 2009) *i.e.* calculating form-factor just for this "averaged position", we see that its phase vanishes. Properly speaking this is some kind of averaging of the scattering amplitude and we obtain for the intensity $I_{av.\ ampl.}$ just the Eq. (7):

$$I_{av.\ ampl.} = |F|^2 \cos^2\left(\pi \frac{L}{\lambda}\right) I_{Laue}(\lambda). \qquad (13)$$

By contrast to this, Eqs. 9 and 10 give in this situation:

$$\Phi(\lambda) = \frac{1}{2}|F|\left(e^{2\pi i a/\lambda} + e^{-2\pi i a/\lambda}\right) = |F|\cos\left(2\pi\,a/\lambda\right), \quad I_{av.\,intens.}(\lambda) = I_{av.\,ampl.}\cos^2\left(\frac{2\pi a}{\lambda}\right) + \dots \quad (14)$$

where we omitted the terms describing the diffuse scattering. Eq. 14 predicts that the intensity of the reflections in $I_{av.\,ampl.}(\lambda)$ (located at the points $d = m\lambda, m = 1, 2, \dots$) should be modulated proportionally to $\cos^2\left(2\pi m a/d\right)$. Particularly, if $a \approx d/4$ the odd reflections corresponding to $m = 1, 3, 5, \dots$ (these are the peaks M3, M9, M15, … in the muscle X-ray terminology) become split and/or extinct. This is quite natural, because now the number of possible positions for MHs doubled ("lattice constant" is twice as little), but they are occupied with a probability of 1/2.

In the event of three discrete states with the form factors

$$F_1 = |F|, \quad F_{2,3} = |F|\exp\left(\mp 2\pi i\,a/\lambda\right) \quad (15)$$

the "averaged position" of the lever arm coinciding with the attachment point (when $p_1 = p, p_{2,3} = (1-p)/2$) is the most interesting again. Now, using the form factor corresponding to this position, we obtain Eq. 13 again, whereas Eqs. 9 and 10 give

$$\Phi(\lambda) = |F|\left[p + (1-p)\cos\left(\frac{2\pi a}{\lambda}\right)\right], \quad I_{av.\,intens.}(\lambda) = \left[p + (1-p)\cos\left(\frac{2\pi a}{\lambda}\right)\right]^2 I_{av.\,ampl.} + \dots \quad (16)$$

Naturally, X-ray pattern varies most dramatically when the multiplier vanishes. For instance, if $p = \frac{1}{2}$ we obtain

$$I_{av.\,intens.}(\lambda) = \left[\cos\left(\frac{\pi a}{\lambda}\right)\right]^4 I_{av.\,ampl.} + \dots \quad (17)$$

In this case, if $a = d/6$ the peaks M9, M15, M21, … become split and/or extinct, etc.

However, these examples do not mean that the position of the lever arm in which it is perpendicular to actin filament always is the most interesting for comparison of Eqs. 7 and 9. The position in which the difference between the diffraction patterns predicted by these formulae becomes maximal should vary from one discrete model to another.

### *Physical content of the equations*

If the conformation of MHs proceeds continuously, then in the experimental conditions, they stay in equal states, and, when the period $d$ is multiple to the wave length, they scatter radiation with an equal amplitude and in one phase. This is the maximally ordered state, and all interference effects in Eq. 7 are connected only with changes in the distance $2a + L$ between the equivalent MHs from different parts of sarcomere.

Let us suppose now that every MH consists of two distinct pieces 1 and 2 with an equal scattering cross-section, which are spaced $d/2$ apart. Then, we obtain a 2$N$-item chain of equal and equidistant (spacing $d/2$) scattering objects, and all the odd reflections $d = (2m+1)\lambda,\ m = 1,2,...$ extinct. On the one hand, we can say that inside every MH two beams scattered by two its pieces cancel out (this corresponds to vanishing of form factor in Eq. 2). On the other hand, in virtue of periodicity, the distance between pieces 1 and 2 of different MHs equals odd number of half-wavelengths. So, we can assert equally well that those very beams that were scattered by different MHs cancel out each other.

Finally, let us suppose that each MH can reside in one of two positions spaced $d/2$ apart and irregularly jumps from one position to another. Further, let at some instant $\alpha N$ MHs stay in the state 1 and $(1-\alpha)N$ - in the state 2. Inasmuch, as the beams scattered by MHs residing in different states cancel out, the multiplier $(1-2\alpha)^2$ should appear in the intensity of any odd peak.

The same mechanism must decrease the intensity of scattered X-ray beam whenever the distance between the scattering centers corresponding to a pair of discrete states equals odd number of half-wavelengths. As a result, Eq. 9 predicts an additional fall of intensities of some reflections as compared to Eq. 7, which is balanced out by the appearance of noncoherent scattering of the same intensity. The magnitude of the effect depends on probabilities of MH residing in these states rather than on the total number of discrete states available for MH. In addition, the effect is preserved in some measure if the distance between scattering centers only mere approximately equals to odd number of half-wavelengths.